\documentclass{elsart}
\usepackage{amssymb,amscd,amsmath,graphicx}
\begin{document}
\runauthor{Goychuk, H\"anggi}
\begin{frontmatter}
%\title{Conformational diffusion model of the ion channel gating}
\title{The role of conformational diffusion 
in ion channel gating}
\author{Igor Goychuk\corauthref{cor1}},
%\author[Augsburg]{  }
\author{ Peter H\"anggi}

%\thanks[X]{}

\address{Universit\"at Augsburg, Institut f\"ur Physik, 
Theoretische Physik I,\\ 
Universit\"atsstr. 1, D-86135 Augsburg, Germany}

\corauth[cor1]{Corresponding author, tel.: +49-821-598-3235,
fax:+49-821-598-3222, e-mail: goychuk@physik.uni-augsburg.de }

\begin{abstract}
We consider an exactly tractable model of the Kramers type for the 
voltage-dependent gating dynamics
of single
ion channels. It is assumed that the gating dynamics is caused
by the thermally activated transitions in a bistable potential. Moreover,
the closed state of the channel is highly degenerate and embraces
the whole manifold of closed substates. Opening of the ion
channel is energetically prohibited from most of the 
closed substates and 
requires a special conformation where the voltage sensor  can move along
an activation pathway and trigger the transition into the open conformation. 
When the corresponding activation barrier towards the channel's
opening is removed by the applied voltage, the statistics of non-conducting
time intervals
become strongly influenced by the conformational diffusion. For the
corresponding supra-threshold voltages, 
our model explains the origin of the power law distribution
of the closed time intervals. 
The
exponential-linear dependence of the opening rate on voltage, 
often used as an experimental
fit, is also reproduced by our model.  
\end{abstract}
\end{frontmatter}

\section{Introduction}

Ion channels present complex protein structures embedded across
the biological cell membranes thereby forming the ion-conducting, 
selective nanopores
\cite{Hille}. The conformational dynamics of these special proteins,
which is known under the label of gating dynamics, 
results in the spontaneous openings and closures of ion channels
\cite{Hille}.  
In the simplest case, this
gating dynamics has the ``all-or-none'' character and can be symbolized
by the following kinetic scheme,  
\begin{eqnarray}\label{scheme}
C \begin{array}{c}  k_o(V) \\[-1.2em] \longrightarrow
\\[-1.3em] \longleftarrow \\[-1.2em] k_c(V) \end{array} O  \;.
\end{eqnarray}
Accordingly, 
the conductance of ion channel 
fluctuates stochastically between
some finite value  and nearly
zero. In the case of voltage-dependent ion channels, like a delayed
rectifier potassium channel, or a Shaker $\rm{K}^{+}$ channel,
the opening rate, $k_o(V)$, and the closing rate, $k_c(V)$, are
both dependent on the applied voltage $V$. 

The invention of the patch clamp technique \cite{Neher} 
enables one to
observe the current and the conductance fluctuations 
in real time with a resolution on  the level of 
{\it single} ion channels.  
Moreover, the study of
the statistics of dwelling time-intervals becomes feasible. 
As a matter of fact, the patch clamp experiments with single ion channels 
pioneered the whole area of single molecular research. 
The gating dynamics of an ion channel with one closed and one
open states, as seen visually in the patch clamp recordings, can be
characterized by the probability densities of closed, $f_c(t)$,
and open, $f_o(t)$ dwelling time-intervals. 
The experimental studies reveal
that in many ion channels the distribution of open dwelling times
is approximately exponential, $f_o(t)=k_c\exp(-k_c t)$. 
However, the distribution of closed time intervals $f_c(t)$ 
frequently involves the sum of many
exponentials, $\lambda_i\exp(-\lambda_i t)$, i.e.,
\begin{equation}\label{fit}
f_c(t)=\sum_{i=1}^{N}c_i\lambda_i\exp(-\lambda_i t),
\end{equation}  
with weights $c_i$,obeying $\sum_{i=1}^{N}c_i=1$.
The rationale behind this fitting procedure is the assumption 
that the closed state
consists of $N$ kinetically distinct 
discrete substates separated by large potential
barriers. The gating dynamics is then captured
by a discrete Markovian kinetic scheme with the rate constants
determined by the Arrhenius law. 

Such an approach presents nowadays the
standard  in the neurophysiology \cite{Colquinon}. 
An essential drawback of it is, however, 
that the number of closed substates needed for
agreement with the experimental data can depend on the range 
of applied voltages 
and temperatures used in the experiment. 
For example, the experimental gating dynamics of 
a Shaker potassium
channel has been reasonably described by a sequential 
8-state Markovian scheme 
with 7 closed
states for a 
temperature about $T=20\;^oC$ \cite{Bezanilla94}. However, 
when the same research  
group extended later on  their model to embrace the small 
range of temperatures $10-20\;^oC$, three additional
closed states have been introduced \cite{Rodriguez98}.
This ``proliferation'' of the number of discrete states, required
for the consistent description of  experimental data, 
is rarely addressed within the approach based on a 
discrete Markovian description.

Moreover, it may happen that for  some channels 
the closed time distribution
can be more conveniently fitted by a stretched exponential 
\cite{lieb87},
or by a power law dependence \cite{millhauser,sansom,add} with a few 
parameters only. 
This observation gave rise to the so-called fractal
models of the ion channel gating dynamics such as put forward, by example, 
by Liebovitch {\it et al.} \cite{lieb87}. The diffusion models introduced
by Millhauser {\it et al.} \cite{millhauser}, L\"auger \cite{lauger} and 
Condat and J\"ackle \cite{condat}
are intrinsically based on the concept of dynamical conformational  
substates in proteins, an idea  which has been pioneered
by Frauenfelder {\it et al.} \cite{frauenfelder}. 
The diffusion models serve as a microscopic justification
for the fractal modeling. On the other hand, the discrete diffusion models
yet present Markovian models with a large number of states. The
non-Markovian, fractal behavior emerges from the projection of a full Markovian
dynamics onto the subspace of two observable states as symbolized by the 
kinetic scheme (\ref{scheme}).

Alternatively, the discrete diffusion can be replaced by the continuous diffusion
on a potential landscape. Then, the distinct minima of this landscape, 
separated by substantial potential barriers, correspond to the 
discrete states in the
mainstream approach. However, it may happen that, depending on the applied
voltage, some of the distinct barriers disappear. Then,
the new features of gating dynamics come into the play, 
which are inevitably missed in the discrete modeling.
It seems therefore that a compromise between the discrete Markovian schemes and
a continuous conformational diffusion approach is 
called for 
\cite{sigg,PNAS}. Especially,
the  continuous 
diffusion models, if set up sufficiently simple to allow an  
analytic treatment, 
are capable to provide a new insight into the problem
of ion channel gating. 
In the present work, we refine and
justify further the approach put forward in Ref. \cite{PNAS}.

\section{Model of gating dynamics}

\begin{figure}
\begin{center}
\includegraphics[width=12cm]{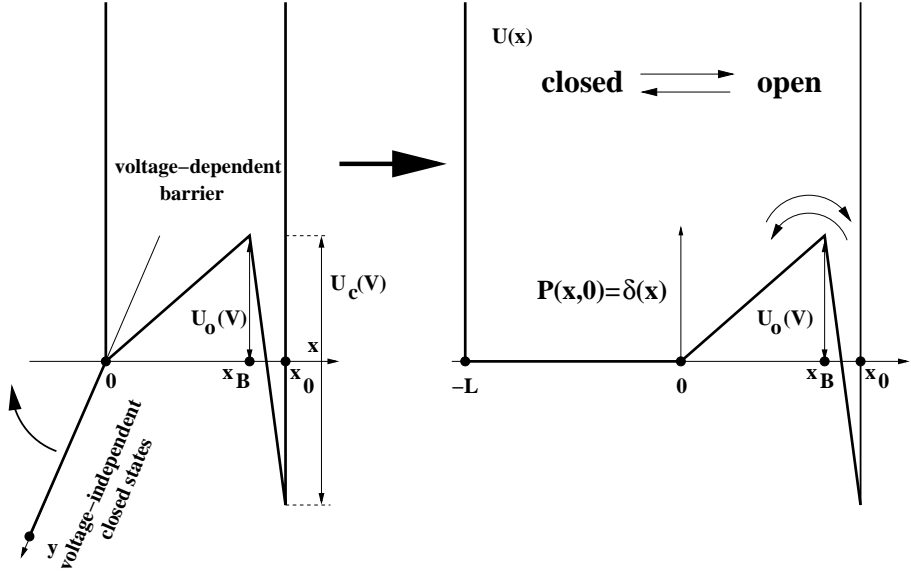}
\end{center}
\caption{Sketch of the studied model. The left  part of figure involves the
two-dimensional conformational space for the gating dynamics. The $x$-axis 
corresponds to the voltage sensor degrees of freedom 
and the $y$-axis to the conformational diffusion.
Note that only the $y=0$ cross-section, $U(x,0)$, of the two-dimensional 
conformational potential $U(x,y)$ possesses a bistable character. 
For $y\geq \delta$ exceeding
some small $\delta$-vicinity of $y=0$, the $y=a=constant$ 
cross-sections, $U(x,a)$, are essentially
monostable, exhibiting sharp minima at $x=0$. Moreover, the potential $U(0,y)$ is
flat in the $y$-direction. These features allow one to simplify the picture of
a two-dimensional reaction state space to the case of one-dimensional
reaction coordinate (see the right part in the figure) as described in the text.}
\end{figure} 

To start, let us consider the model depicted in Fig. 1. Its central element
is the voltage-dependent transitions in a bistable, piecewise
linear potential  along
the $x$-axis, see l.h.s. in Fig. 1. This bistable potential
corresponds to the motion of the so-called voltage sensor. The voltage
sensor presents a
functional part of the ion channel machinery which changes 
its position in response
to the changes in the transmembrane potential. 
In accord with the current view \cite{Hille}, the voltage sensor is 
formed by the system of four
positively charged S4 $\alpha$-helices which carry a total
gating charge $q\sim 10-13\e_0$, where $e_0$ is the  
positively-valued, elementary charge. 
When $x=0$, the voltage sensor is in its resting position; when
$x=x_0$, the voltage sensor is fully activated and the complete gating charge
is moved across the membrane. Moreover,  
the drop $V$ of the transmembrane
electric potential is assumed to occur linearly on the characteristic 
length $x_0$ which
corresponds to the {\it effective} width of the cellular membrane
within the region of the gating charge translocation. Then,
the energy barriers, $U_{o}(V)$ and $U_{c}(V)$, follow as   
\begin{eqnarray}
U_o(V) & = &qr(V_c-V),\label{Uo}\\
U_c(V) & = &U_c^{(0)}-q(1-r)(V_c-V) \label{Uc},
\end{eqnarray}
where $r:=x_B/x_0$.  Furthermore, $V_c$
in (\ref{Uo}) corresponds to the threshold value of the voltage $V$ when the 
activation
barrier towards the channel opening vanishes and $U_c^{(0)}$ in  
(\ref{Uc}) is the corresponding height of the activation barrier towards
the channel closing. Note that for $V>V_c$, $U_o(V)$ is negative, but
$U_c(V)$ has always a large positive value, since one assumes
that $U_c^{(0)}$ is large
and $1-r<<1$. These assumptions correspond to the experimental
observations that the closing rate has a very strong dependence
on the temperature and it is less voltage-sensitive than the opening rate
\cite{Hille}.
Furthermore, one assumes that
the voltage sensor located at the position $x>x_B$ ($x_B$ corresponds to the
top of the 
activation barrier  for the sub-threshold voltages, $V<V_c$) triggers the
conformation change in the activation gate. This latter 
conformational change then 
leads to the
channel opening. It is worth to notice that in accord with the 
reasoning in \cite{Mac1,Yellen},
the activation gate is likely formed by the bundle of the intracellular ends of
inner, pore-lining S6 $\alpha$-helices \cite{Hille}.
The motion of the voltage sensor creates an indirect mechanical torque on
the activation gate. This torque is mediated by other structural elements
of the channel protein. For this reason, the link between the voltage sensor and the 
activation gate may be flexible and this circumstance can introduce 
a kind of mechanical hysteresis. Namely, the closure of the activation
gate does not occur immediately, when the voltage sensor crosses the border
at $x=x_B$ in the back direction, but rather requires that (is most probable when)
the voltage sensor is fully returned to its resting position at $x=0$.
For this reason, by calculating the dwelling time distribution $f_c(t)$
we assume the initial condition for the 
probability density $P(x,t)$ in the form $P(x,0)=\delta(x)$, i.e.,
each and every closing time-interval starts when the voltage sensor has
fully returned. This inevitably presents an approximation to the reality.
 In accord with the general activation theory of Kramers  
\cite{HTB90}, we also
have to impose an absorbing boundary at $x>x_B$. The most natural choice
for this is $x=x_0$. 
However, in view of our model assumption
$U_c^{(0)}\gg k_B T$ one can safely move the absorbing boundary to $x=x_B$,
i.e., $P(x_B,t)=0$ at all times $t\geq 0$.
 The reason for this is that the final transition $x_B\to x_0$ leading
 to the opening
is very quick. The same is assumed in many discrete 
Markovian schemes \cite{Hille}. Given this assumption, 
the duration of this final
step is negligible in calculating $f_c(t)$. Furthermore,
in order to find the distribution  of open times, $f_o(t)$, one should put
the reflecting boundary at $x=x_0$ and the sink (absorbing
boundary) at $x=0$, in accord with the above discussion. 

%One should bear
%in mind however that the {\it effective} gating charge, which senses the
%transmembrane potential, for the backward, closing
%transition can be different (in fact, much less) than the gating charge $q$
%of the closed channel. 
%The reason for this is the
%following. When the channel is closed, its wide (about 12 \AA $\;$ in diameter)
%internal cavity \cite{Hille} is filled up with water. When it opens, this
%cavity, as well as the selectivity filter, occasionally fill 
%in with the {\it positively} charged permeable ions
%${\rm K}^{+}$. The presence of several extra charges inside the channel
%can essentially influence the energy barriers of the voltage sensor and
%destabilize its open position what can be accounted for by the gating
%charge reduction.

The next important element of our model is the assumption that
the voltage sensor can operate only if the protein acquires a special
pre-open conformation. Otherwise, its motion is
energetically prohibited due to the steric factors. To
account for this type of behavior we introduce 
the additional, ``perpendicular'' conformational reaction 
coordinate $y$, see l.h.s. in
Fig. 1. It describes the conformational diffusion.
One assumes that the corresponding dynamics is only weakly voltage-dependent
and we shall neglect the accompanying redistribution of charges in the
protein molecule.
Actually, this conformational diffusion should occur in a 
hierarchically arranged, ultrametric conformational space 
\cite{frauenfelder,dobson,metzler}. 
In a simplified manner, one can model
this hierarchical features by a random walk with the mean-squared 
displacement
$\langle \Delta y^2(t)\rangle \propto t^{\gamma}$. This latter
relation holds, strictly speaking, if the diffusion is
unbounded in space. The bounded character, however, is important for the 
following.
We restrict ourselves in this paper to the case of normal diffusion, 
$\gamma=1$, as the simplest
possibility. Finally, within
this outlined picture one can ``convert'' the dynamics along
the $y$-axis into the dynamics along the $x$-axis, by use of the 
extra part $[-L,0]$ there, 
see r.h.s in Fig. 1,  
whereby introducing the auxiliary diffusion length $L$ and the diffusion
coefficient $D=k_BT/\eta$.  Here, $\eta$ is the viscous friction
coefficient characterizing the conformational dynamics. 
In order to account for the bounded character of the
conformational diffusion, we impose the 
reflecting boundary condition
$\frac{\partial P(x,t)}{\partial x}|_{x=-L}=0$
at $x=-L$ for all $t$. 
Note that $\eta$ should generally depend exponentially on the temperature, i.e.
\[
\eta=\eta_0\exp(\epsilon/k_BT),\] 
where $\epsilon$ is 
a characteristic activation energy for transitions between conformational
microstates. It can be in the range of several $k_BT_{room}$. 
The parameters $D$ and $L$ are conveniently combined into the single parameter --
the conformational diffusion time, i.e.,
\begin{eqnarray} \label{tauD}
\tau_D=L^2/D \propto \exp(\epsilon/k_BT)/k_BT,
\end{eqnarray}
which constitutes an important parameter of the theory.

\section{Theory of ion channel gating}
 
In order to find $f_c(t)$ one has to solve the Smoluchowski dynamics
\begin{eqnarray}\label{kramers}
\frac{\partial P(x,t)}{\partial t}=D\frac{\partial}{\partial x}
\left(\frac{\partial}{\partial x}+\beta\frac{\partial U(x)}{\partial x}
\right) P(x,t),
\end{eqnarray}
where $\beta=1/(k_BT)$, supplemented by the initial and the boundary 
conditions
of reflection and absorption discussed above. 
The closed residence time distribution then follows as
\begin{equation}\label{def1}
f_c(t)=-\frac{d\Phi_c(t)}{dt},
\end{equation}
where $\Phi_c(t)=\int_{-L}^{x_B}P(x,t)dx$ 
is the survival probability  in  the closed state.

By use of the standard Laplace transform method 
we arrive at the following  
{\it exact} solution for the Laplace-transformed distribution
of closed times $\tilde f_c(s)$, see in Ref. \cite{PNAS}:
\begin{equation}\label{result4}
\tilde f_c(s)=\frac{A(s)}{B(s)},
\end{equation}
where 
\begin{eqnarray}\label{result5}
A(s) &= & 
\exp(-\beta U_o(V)/2)\sqrt{\beta^2 U_o^2(V)+4\xi^2\tau_D s} \\ \label{result6}
B(s)& = &\sqrt{\beta^2 U_o^2(V)+4\xi^2\tau_D s}\cosh\Big( \sqrt{
\beta^2 U_o^2(V)+4\xi^2\tau_D s}/2\Big)\\ \nonumber
&+&\Big(2\xi\sqrt{\tau_D s}
\tanh\sqrt{\tau_D s}-\beta U_o(V) \Big) \sinh\Big( \sqrt{
\beta^2 U_o^2(V)+4\xi^2\tau_D s}/2\Big), 
\end{eqnarray}
where the parameter $\xi$ is given by $\xi:=x_B/L$.
 The explicit result in Eqs. (\ref{result4})-(\ref{result6}) allows one to find 
all moments of
the closed residence time distribution. In particular, the
mean closed residence time, 
\[\langle T_c\rangle:=\int_{0}^{\infty}
tf_c(t)dt =\lim_{s\rightarrow 0}
[1-\tilde f_c(s)]/s,\]
emerges as 
\begin{eqnarray}\label{result1}
\langle T_c(V)\rangle=\tau_D\xi
\frac{\beta U_o(V)(e^{\beta U_o(V)}-1-\xi)+\xi(e^{\beta U_o(V)}-1)}
{\beta^2 U_o^2(V)}.
\end{eqnarray}
The effective opening rate can be defined as 
$k_o(V):=1/\langle T_c(V)\rangle$. Let us consider the limiting
case $\xi=x_B/L\ll 1$. In the language of discrete substates, this limit 
is tantamount to the assumption that the number of quasi-degenerate
conformational substates, which correspond to the resting position
of the voltage sensor, largely exceeds that of the voltage-sensor.
Under this assumption, we obtain in leading order of $\xi$ 
\begin{equation}\label{result7}
k_o(V)=\frac{1}{\langle T_c \rangle}
\approx\frac{1}{\xi\tau_D}\frac{\beta rq(V-V_c)}
{1-\exp[-\beta rq (V-V_c)]}\;.
\end{equation}
Note that the functional form in Eq. (\ref{result7}) is nothing but
the familiar exponential-linear dependence used as a
phenomenological experimental fit
in the celebrated paper by Hodgkin and Huxley \cite{Hodgkin} used
to describe the voltage-dependence of the opening rate of a single gate
in the potassium channels. This form is commonly used to 
parameterize the opening
rate of the potassium channels, see, e. g., in \cite{Braun}.
Our model provides a detailed justification for this experimental result.
Its remarkable feature is that the dependence of the rate on voltage
is exponential for $V<V_c$, when the energy barrier towards
activation of the voltage sensor is essential, $k_BT < U_o(V)$. 
This exponential 
voltage-dependence implies in virtue of (\ref{tauD})  
an exponential dependence
on temperature as well, i.e.,
\begin{equation}\label{temper1}
 k_o\propto \exp\{-[\epsilon+rq(V_c-V)]/k_BT\}\;. 
\end{equation}
This exponential temperature dependence has two sources: 
a first one is due to the activation barrier of the voltage 
sensor $U_o(V)$, while a second one is due to the activation 
barrier $\epsilon$ 
between diffusional 
microstates, which we have assumed for reasons of  simplicity to be of equal
height. The barrier $U_o(V)$ can acquire large values.
For example, assuming typical values $V_c=-40$ meV, 
$rq=10\;e_0$ and a room temperature $k_BT_{room}=25\;{\rm meV}$ one obtains
$U_o(V_r)=20\;k_BT$ for the cell resting potential $V_r=-90\;{\rm meV}$.
Furthermore, when the activation barrier $U_o(V)$ vanishes for $V>V_c$,
it follows from (\ref{result7}) that the rate $k_o(V)$ exhibits the linear
dependence on voltage, i.e., 
$k_o(V)\propto (V-V_c)$. In this case, its 
temperature dependence is distinctly reduced and becomes mainly determined  
by the activation barrier $\epsilon$ of the conformational
diffusion.  The latter one can assume a few $k_BT_{room}$ only. 
The very different 
temperature dependences of the opening rate for $V\ll V_c$ and for $V>V_c$
present an interesting feature of our model which calls for an
experimental verification.

\begin{figure}
\begin{center}
\includegraphics[width=10cm]{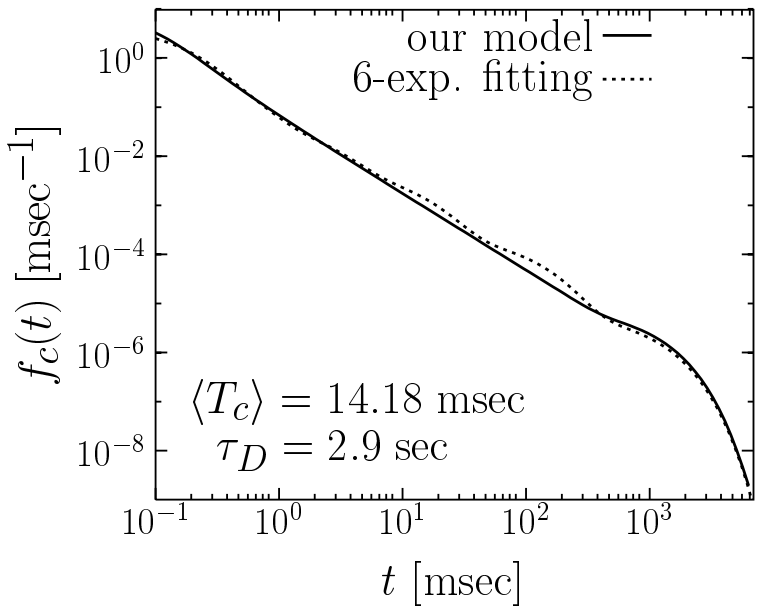}
\caption{Closed time probability density from our model, Eq. (7), (solid line)
 and 6-exponential
fitting procedure of the experimental data from Ref. \cite{sansom} 
(dotted line). The following
set of parameters is used in our calculations: $\tau_D=2.9\;{\rm sec}$, 
$\xi=0.01$ and $\beta U_o(V)=-1.653$.}
\end{center}
\end{figure}

\begin{figure}
\begin{center}
\includegraphics[width=12cm]{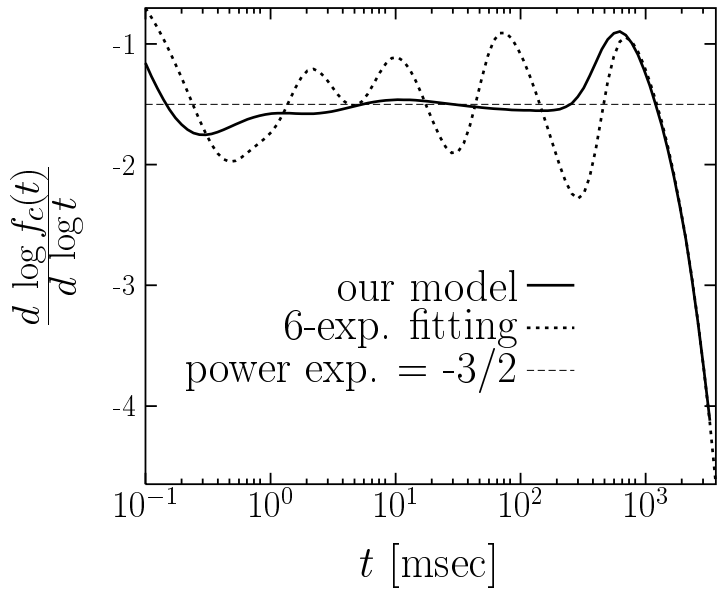}
\caption{Logarithmic derivative of the closed time probability
density $f_c(t)$.
The solid line and the dotted line correspond to those in Fig. 2; the 
long-dashed level line corresponds to 
 strict power law dependence $f_c(t)\propto t^{-3/2}$.
 The oscillating behavior reveals the hierarchical character of
 the conformational dynamics.}
\end{center}
\end{figure}

It is worthwhile to notice that the discussed crossover from an
exponential to linear
 voltage dependence of the opening rate  is qualitatively preserved for
any value  of $\xi$, including the extreme case $\xi\gg 1$. In 
this case $L\ll x_B$ and
the conformational diffusion does not play a dominant role.  The discussed 
feature is caused by the vanishing of the 
activation barrier $U_o(V)$ for $V>V_c$. However, the distribution of
closed time-intervals $f_c(t)$ depends {\it qualitatively} on $\xi$.
Namely, for $\xi\ll 1$ and $V>V_c$ it displays a power law regime, i.e.
\begin{equation}
f_c(t)\approx \frac{1}{2(\pi\tau_D)^{1/2}k_ot^{3/2}}, 
\end{equation}
 for
the intermediate time scale 
$\langle T_c\rangle^2/\tau_D\ll t\ll \tau_D$. 
In Fig. 2, we illustrate
this prominent feature for the following 
model parameters: $\tau_D=2.9\;{\rm sec}$, $\xi=0.01$ and $\beta U_o(V)=
-1.653$. The solid line denotes the closed time
probability density obtained from our model, Eqs. 
(\ref{result4})-(\ref{result6}), by a numerical inversion of the Laplace
transform $\tilde f_c(s)$. The short-dashed line presents
the fitting of  the experimental data for the delayed rectifier
${\rm K}^+$ channel from a neuroblastoma $\times$ glioma cell by use of Eq. 
(\ref{fit}) with 6 exponentials. This
fitting procedure is taken
from Ref. \cite{sansom} (see, Table 4 therein) and implicitly corresponds to
a discrete Markovian scheme with 6 closed substates. Both results
describe well the intermediate power law trend and the exponential tail
of the closed time-interval distribution. Nevertheless, some
small distinctions can be distinguished. 

%%%%%%%%%%%%%%%%%%%%%%%%%%%%%%%%%%%%%%%%%%%%%%%%%%%%%

The plot of the logarithmic derivative, 
 $d \log f_c(t)/d \log t$, versus the logarithm of time $t$ in Fig. 3 
 renders  these distinctions 
 much more visible.  The advantage of
 such a representation as in Fig. 3 is -- 
 in accordance with the reasoning in Ref. \cite{metzler} -- that 
 the hierarchical, tree-like 
 relaxation
 dynamics reveals itself by  logarithmic oscillations around the
 level line that corresponds to the power law trend. Remarkably enough, the
 exponential fit from Ref. \cite{sansom} does exhibit such
 logarithmic oscillations, cf. Fig. 3. Thus, this result seems to 
 support the hierarchical character of the conformational dynamics.
 Our simplified model does not distinctly display
 these fine features
 as these are rooted in the discrete nature of hierarchical states. 
 Nevertheless,
 the power law trend, which reveals the presence and
 the importance of the conformational dynamics, is reproduced by our model. 
 Moreover,
 its appealing feature is that it requires only few parameters
 which possess a clear physical meaning.  The
 particular value of the power law exponent $-3/2$ corresponds to
 the  conformational dynamics  
 modeled as a bounded normal diffusion.
 Other power law exponents, also seen experimentally \cite{add},
  require a generalization of our model to the case of 
 anomalous diffusion. Such corresponding work is presently in progress.

\section{Summary and conclusions}

We herewith have presented a simple model of the complex gating dynamics in
voltage-dependent potassium ion channels. It is based on the
concept of  conformational diffusion. In particular, we assumed that the
conformational change leading to the opening of ion channel is
triggered by  the voltage sensor which, in its turn, is
permitted only when the channel  protein acquires a special pre-open
configuration. When the ion channel is closed, it 
exhibits an internal, conformational diffusion over the manifold of
conformational substates which do not possess a sensitive voltage
dependence.  In a simplified manner, this conformational diffusion
has been mathematically modeled by bounded normal diffusion.
Moreover, it has been assumed that the open conformation of the channel
is more ordered, with less conformational substates.
Then, the conformational diffusion does not play an essential  role. In the
language of statistical thermodynamics this means that  the
ion channel upon opening undergoes a kind of ordering transition 
into a state with
lower configurational entropy. 

We should also stress here that our simple
model is aimed not to replace the standard discrete Markovian 
modeling \cite{Colquinon},
but rather to complement these efforts by highlighting some basic physical 
features which otherwise become blurred with the standard approach.
In particular,
it has been shown that the transition from an exponential to linear voltage
dependence of the opening rate occurs when the activation barrier for the
voltage sensor towards the channel's opening vanishes due to the applied 
transmembrane voltage. Moreover, if the
conformational diffusion time $\tau_D$ exceeds largely the mean
duration of closed time intervals $\langle T_c\rangle$, 
the closed time distribution exhibits
a power law feature on the intermediate time scale 
$\langle T_c\rangle ^2/\tau_D\ll t\ll \tau_D$. This power law changes over
into an exponential tail for times $t>\tau_D$. These features are
seemingly
consistent with the experimental observations for some ${\rm K}^{+}$-channels.  
The true physical benchmark of our model is, however, the prediction 
that the opening 
rate will become
much less temperature-dependent for supra-threshold voltages.
The weak temperature dependence in this latter regime should correlate
with a weak voltage dependence. This distinct prediction 
calls for an experimental verification, which hopefully will support
our reasoning for the gating dynamics in ion channels.

{\bf Acknowledgments}. This work has been supported by the Deutsche
Forschungsgemeinschaft via the Sonderforschungsbereich SFB-486,
{\em Manipulation of matter on the nanoscale}.

%\appendix
%\section{APPENDIX}

\end{document}